\documentclass[journal]{IEEEtran}
\usepackage{color}
\usepackage{graphicx}
\usepackage{url}
\usepackage{multirow}
\usepackage{here}
\usepackage{fancyhdr}

\ifCLASSINFOpdf
\else
\fi
\graphicspath{ {./} }

\begin{document}
%
\title{Tomographic Imaging by a Si/CdTe Compton Camera for $^{111}$In and $^{131}$I Radionuclides}
%
%
%

\author{Goro~Yabu,
  Hiroki~Yoneda,
  Tadashi~Orita,
  Shin'ichiro~Takeda,
  Pietro~Caradonna,
  Tadayuki~Takahashi,
  Shin~Watanabe,
  Fumiki~Moriyama%
  \thanks{G.Yabu is a member of Department of Physics, The University of Tokyo, Kavli Institute for the Physics and Mathematics of the Universe (WPI), 5-1-5 Kashiwa-no-ha Kashiwa Chiba 277-8583, Japan.}
  \thanks{T.Orita, S.Takeda, T.Takahashi are members of Kavli Institute for the Physics and Mathematics of the Universe (WPI), Department of Physics, The University of Tokyo, 5-1-5 Kashiwa-no-ha Kashiwa Chiba 277-8583, Japan.}
  \thanks{H.Yoneda is a member of RIKEN Nishina Center for Accelerator-Based Science (RNC), RIKEN, Hirosawa 2-1, Wako, Saitama, 351-0198, Japan.}
  \thanks{T. Orita and S. Takeda are also with iMAGINE-X Inc., 5-38-6 Jingumae, Shibuya-ku, Tokyo 150-0001, Japan (E-mail: shinichiro.takeda@imagine-x.jp).}
  \thanks{F.Moriyama is a member of Okinawa Institute of Science and Technology Graduate University, 1919-1 Tancha Onna-son Okinawa 904-0495, Japan.}
  \thanks{S.Watanabe is a member of Institute of Space and Astronautical Science, Japan Aerospace Exploration Agency, 3-1-1 Yoshinodai, Chuo Sagamihara, Kanagawa, 252-5210, Japan}
}


\maketitle

\begin{abstract}
    Tomographic imaging with radionuclides commonly used in nuclear medicine, such as $^{111}$In (171 and 245~keV) and $^{131}$I (364~keV), is in high demand for medical applications and small animal imaging.
    The Si/CdTe Compton camera with its high angular and high energy resolutions is an especially promising detector to extend the energy coverage for imaging to the range that covers gamma-ray emitted from these radionuclides.
    Here, we take the first steps towards short-distance imaging by conducting experiments using three-dimensional phantoms composed of multiple sphere-like solutions of $^{111}$In and $^{131}$I with a diameter of 2.7~mm, placed at a distance of 41~mm.
    Using simple back-projection methods, the positions of the sources are reproduced with a spatial resolution of 11.5~mm and 9.0~mm (FWHM) for $^{111}$In and $^{131}$I, respectively.
    We found that a LM-MLEM method gives a better resolution of 4.0~mm and 2.7~mm (FWHM).
    We resolve source positions of a tetrahedron structure with a source-to-source separation of 28~mm.
    These findings demonstrate that Compton Cameras have the potential of close-distance imaging of radioisotopes distributions in the energy range below 400~keV.
\end{abstract}


\IEEEpeerreviewmaketitle

\thispagestyle{fancy}
\pagestyle{fancy}
\chead[]{
    \fontsize{6}{8}\selectfont \copyright 2021 IEEE.  Personal use of this material is permitted.  Permission from IEEE must be obtained for all other uses, in any current or future media, including reprinting/republishing this material for advertising or promotional purposes, creating new collective works, for resale or redistribution to servers or lists, or reuse of any copyrighted component of this work in other works.
}
\renewcommand{\headrulewidth}{0.0pt}

\section{Introduction}

\label{sec:intro}

The Compton camera is a type of gamma-ray detector which utilizes Compton scattering in the detector.
The direction of an incident gamma-ray is determined by solving the Compton scattering equation.
It is regarded as one of the most promising  detectors to  observe gamma rays with energies from several tens of keV to several MeV, where Compton interaction is the dominant process of photon interactions.
Unlike most other types of gamma-ray detectors, the Compton camera requires no collimators to determine the directions of 
an incident gamma-ray.
Therefore, it has the potential to achieve a high detection efficiency and a wide field of view at the same time.
One of the promising applications of 
Compton cameras is in medical imaging, which achieves this by means of detecting gamma-rays emitted from radionuclides \cite{Suzuki2013,Kataoka2018a,Nakano2019}.

A key detector parameter for medical imaging is the angular resolution, because it determines the accuracy of the location of gamma-ray emitting sources.
Moreover, since the angular resolution also plays a factor in the amount of background contamination, it is important to have high angular resolution to increase the signal to background ratio.
The angular resolution of a Compton camera is determined from the energy and positional resolutions of the detector.
In addition, the resolution is also limited by the finite momentum distribution of the bound atomic electrons in the detector material of the scattering part of the camera.
This is called the Doppler-broadening effect  and is a dominant factor affecting the angular resolution of a Compton camera below a few hundred keV\cite{Zoglauer2003,TakedaPhD}.

In using Compton cameras to observe commonly used radionuclides, such as $^{111}$In and $^{131}$I, that emit low-energy gamma-rays below $\sim$400~keV and widely used in pre-clinical and clinical imaging, we have to ensure a sufficient angular resolution comparable with (or better than) those of the existing detectors, notably SPECT and PET.
For this purpose, we have been developing multi-layered Si/CdTe Compton cameras (Si/CdTe CCs) \cite{Takahashi2004, Watanabe2005,Takeda2007}.
Owing to their good performance in energy  and  positional  resolutions, and the use of Si as the scattering material, which reduces the Doppler broadening effect, the Si/CdTe CCs have achieved a high angular resolution and provide high sensitivity of detecting sources \cite{Takahashi2004, Watanabe2005, TakedaPhD, Takahashi2012}.
In previous results, we obtained an angular resolution of 17~degrees FWHM (full width at half maximum) at 81~keV and 10~degrees FWHM at 122~keV. 
The angular resolution improves for higher incident energies, such as at 511~keV where the resolution is measured to be 2.5 degrees FWHM \cite{Takeda2007}.
Si/CdTe CCs have a typical energy resolution of $\sim$4-5~keV (FWHM), which allows us to resolve multiple lines emitted from sample containing a mixture of radioisotopes \cite{Takeda2007}.

In our previous work, we reported the results of 3D imaging of a mouse using a Si/CdTe Compton Camera in 2013 \cite{Suzuki2013}.
Recently, Kataoka et al. (2018) \cite{Kataoka2018a} presented their results of tomographic imaging of mice obtained with a GAGG-scintillator-based Compton Camera.
However, these experiments focused on gamma-rays with energies above 300~keV.
In order to utilize a Compton camera in an energy range that covers both 171~keV and 245~keV for the two major gamma-ray line emissions emitted from $^{111}$In, we need to investigate the performance at these low energy range.
When used in small animal imaging, e.g., mice and rats, three-dimensional imaging capability at short distances ($\sim$ few cm) is crucial, and hence evaluation of the performance is also one of the focal points of this work.

The primarily motivation for developing methods that can extract medical information from the output data generated by Compton camera systems is driven by the potential of developing novel medical imaging systems.
The list-mode maximum-likelihood expectation maximization (LM-MLEM)\cite{Wilderman1998,Harrison2004} is the often used method to reconstruct images from event-based (list mode) data.
However, previous LM-MLEM based experiments were restricted to imaging using high energy gamma rays or limited to imaging sources located at much greater distances compared to the imaging distances considered in this work.
Also, most of the past simulation studies of imaging algorithms assumed a simple detector response function \cite{Cree1994,Parra2000,Tomitani2002}.

In this paper, 
we measure the performance of a Si/CdTe Compton camera at short-distance tomographic imaging of radionuclide targets in the energy range of 171~keV to 364~keV.
In Section \ref{sec:analysis}, we provide experimental results using several point source radioisotopes and measure some fundamental properties of the camera such as the energy and angular resolutions.
In Section \ref{sec:design}, we perform tomographic imaging, using a 3D phantom of $^{111}$In and $^{131}$I, and evaluate the spatial resolution of the targets placed at a distance of $\sim$41~mm.
Finally, we conclude our result in Section \ref{sec:result}.

\section{Si/CdTe Compton camera}\label{seq:si_cdte}

\nopagebreak
\IEEEPARstart{T}{he} Si/CdTe CC used in our experiments consists of 2 scattering layers made of silicon double-sided strip detectors (Si-DSDs) and 3 absorption layers made of CdTe double-sided strip detectors (CdTe-DSDs).
These detectors are stacked with a pitch of 4~mm.
Table~\ref{tab:spec} summarizes the specifications of the detectors.
Each layer has a detection area of 32~mm $\times$ 32~mm with a positional resolution of 250~$\mu$m.
Each Si-DSDs and CdTe-DSDs are 500~$\mu$m and 750~$\mu$m thick, respectively.
The CdTe-DSD has an Al/CdTe/Pt configuration, which is based on the high-resolution CdTe-diode detector technology~\cite{Takahashi1998}.
The detectors were operated at a temperature of $-$20$^{\circ}$C, and a bias voltage of 250~V was applied.

The low-noise analog ASICs of the Si-DSD and CdTe-DSDs digitize the pulse heights in reading the signals from each strip \cite{Watanabe2014}.
To begin with, the triggers from the Si layers and the CdTe layers are ORed independently and become triggers at the scattering part (Si) and the absorbing part (CdTe).
When two triggers are detected within a 600~nanoseconds time window, the signals from all layers are taken as a candidate of Compton scattering events in the camera.
If the signals are spread over adjacent strips, those data are merged to extract the position and the deposited energy, corresponds to each interaction.
Further information about the CdTe-DSDs and its data processing are found in~\cite{Watanabe2009, Ishikawa2010}.
Hereafter, the reconstructed energy and position, which corresponds to the data of one interaction, is referred as a hit.

As summarized in Table~\ref{tab:spec}, the measured FWHM energy resolutions for a
Si-DSD and CdTe-DSD are 5.5$\%$ at 31~keV and 1.9$\%$ at 356~keV, respectively.
In practice, the energy is determined as the sum of the energies of a signal hits from a Si-DSD and a CdTe-DSD.
The resolution is found to be 4.1~keV at 171~keV and 5.0~keV at 245~keV. The maximum counting rate
of the system is about 1~kcps.

Silicon is used as the scattering medium in Compton cameras because Doppler-broadening effect, which degrades the angular resolution, is relatively small compared to other material.
Appendix \ref{appen:doppler} summarizes our study of the Doppler-broadening effects in silicon and we considered a detector with a different scattering medium besides silicon.

We place the front detector of the camera at a distance of 41 mm away from the vertical axis of the target stage (see Fig.\ref{fig:setup}).
The target is mounted on a rotary stage and measurements of the target can be performed from arbitrary angles.

\section{Analysis method}\label{sec:analysis}

\nopagebreak
\IEEEPARstart{B}{efore} conducting tomographic imaging, we investigated the angular and spatial resolutions of the Si/CdTe CC, comparing the experimental results with  Monte Carlo simulations for a single point source.

The angular resolution of the Compton camera is defined as the FWHM of the Angular Resolution Measure (ARM) \cite{Zoglauer2003}, which is the difference between the scattering angles, $\Delta \theta \equiv \theta_{K} - \theta_{G}$, where $\theta_{K}$ is calculated using the energies of the scattered electron and scattered photon and $\theta_{G}$ is calculated using the measured momentum directions of the incident and scattered photons.

To evaluate the angular resolution of the Si/CdTe CC for various energies, we conducted experiments in each of which one of $^{133}$Ba, $^{22}$Na point sources and a 10~$\mu$L droplet of $^{111}$In were placed at a distance of 41~mm from the detector.
The procedure of obtaining the ARM distributions is described in the following sub-sections.

\begin{table}[htb]
    \centering
    \caption{Specifications of our Si/CdTe semiconductor Compton camera}
    \begin{tabular}{ccc} \hline
        Detector           & Si-DSD                  & CdTe-DSD \\ \hline \hline
        \multirow{2}{*}{\# of layers} & \multirow{2}{*}{2} & \multirow{2}{*}{3} \\
        & & \\ \hline
        \multicolumn{1}{c}{\multirow{2}{*}{Detecting area}} & \multicolumn{2}{c}{\multirow{2}{*}{32$\times$32~mm$^2$ }} \\
        & & \\ \hline
        \multicolumn{1}{c}{\multirow{2}{*}{Thickness}}         & \multirow{2}{*}{500~$\mu$m} & \multirow{2}{*}{750~$\mu$m} \\
        & & \\ \hline
        \multicolumn{1}{c}{\multirow{2}{*}{Spatial resolution}} & \multicolumn{2}{c}{\multirow{2}{*}{250~$\mu$m}} \\
        & & \\ \hline
        \multirow{4}{*}{ \begin{tabular}{@{}c@{}} Energy resolution \\ (FWHM) \\ \end{tabular} } &
        \multirow{4}{*}{5.5$\%$ @ 31~keV} &
        \multirow{4}{*}{ \begin{tabular}{@{}c@{}} 2.8$\%$ @ 171~keV \\ 1.9$\%$ @ 356~keV \\ 1.2$\%$ @ 511~keV \end{tabular} } \\
        & & \\
        & & \\
        & & \\
        \hline
    \end{tabular}
    \label{tab:spec}
\end{table}

\subsection{Analysis to obtain the ARM distribution}\label{sec:selection}

Gamma rays in our expected energy range undergo either or both of (potentially multiple) Compton scattering and photo absorption in the detector and information of the energy and position for both  processes are required to calculate the ARM.
Low energy sums were considered as noise and were thus filtered out. The energy thresholds for filtering were set to 5 and 10 keV for Si-DSDs and CdTe-DSDs, respectively.
For energies below 0.5 MeV, a photon will likely Compton scatter once through the silicon scattering plate used here. 

\begin{figure}[htb]
    \centering
    \includegraphics[width=0.9\linewidth]{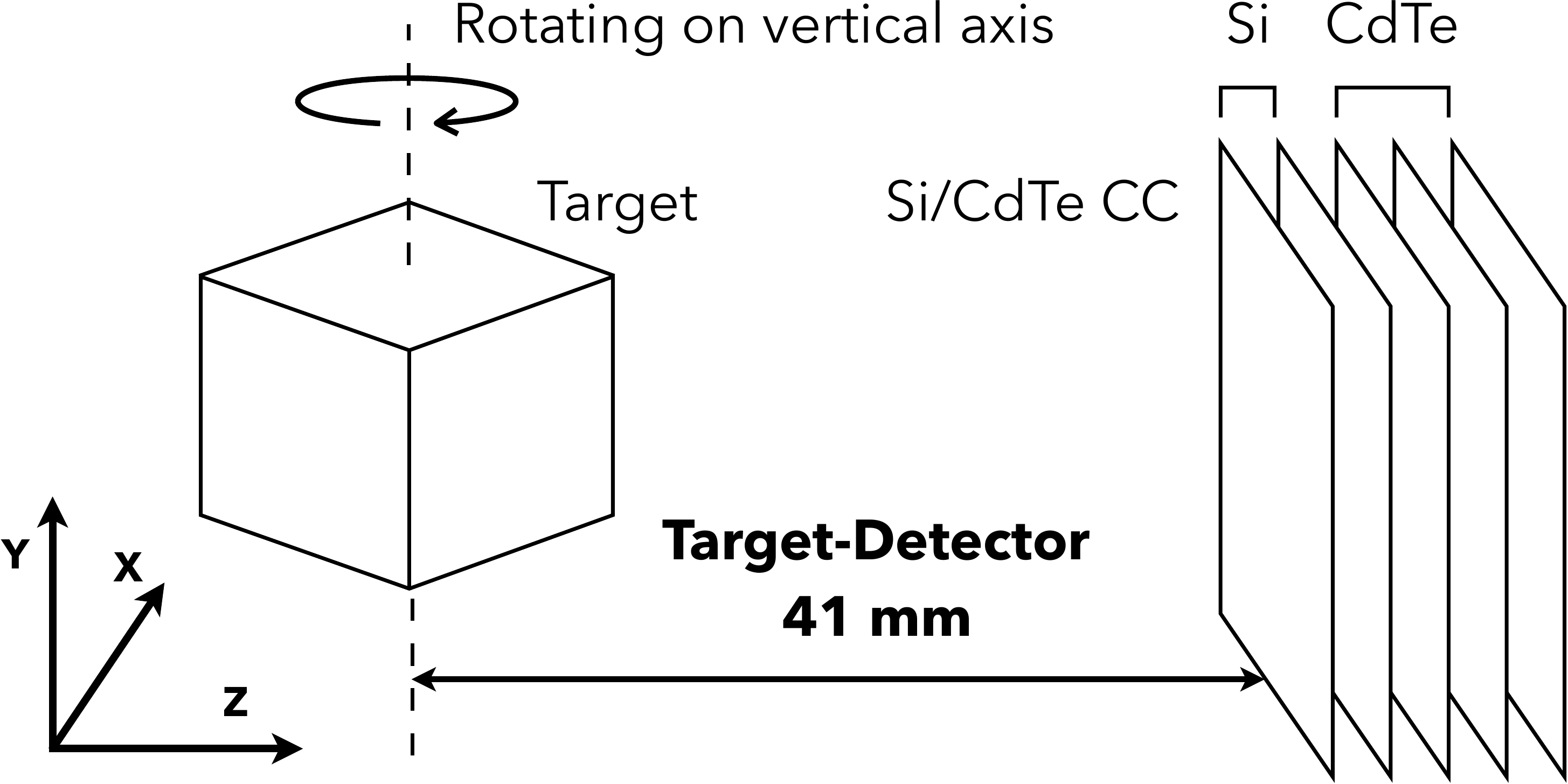}
    \caption
    {
    Configuration of the experiment system.
    The distance between the target and the Si/CdTe CC front detector is 41~mm.
    The stacking pitch of the detectors is 4~mm.
    The target can be rotated  on a rotary stage and measured from an arbitrary angle of view.
    The axes of $x$,$y$, and $z$ of the coordinates are indicated.
    The $y$-axis corresponds to the vertical axis.
    }
    \label{fig:setup}
\end{figure}

\begin{figure*}[htb]
    \centering
    \includegraphics[width=\linewidth]{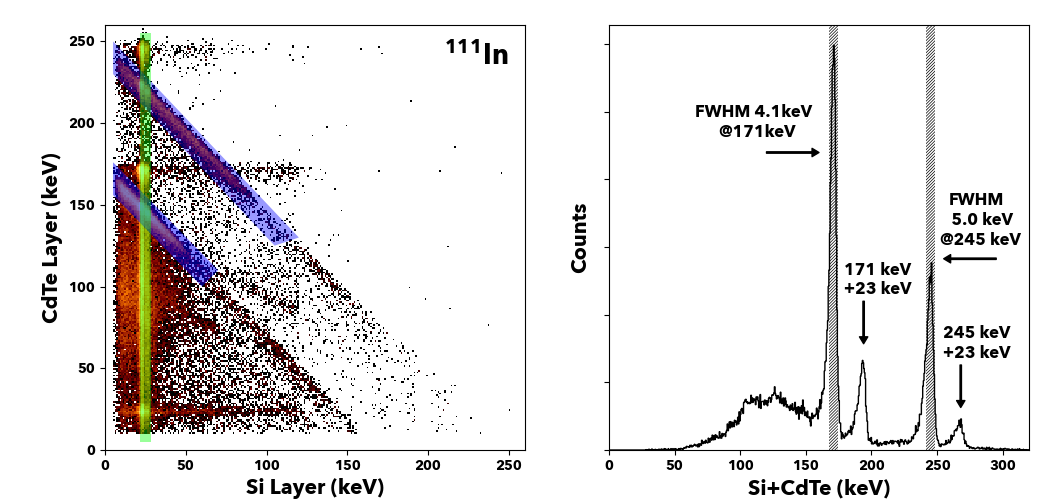}
    \caption
    {
    (Left) Scatter plot of the energies of the hits at  the Si-DSDs and  CdTe-DSDs for $^{111}$In experimental data.
    The regions shaded in green shows the Si-DSD energy range of 23--27~keV, which corresponds to the fluorescence K emission lines.
    (Right) Energy spectrum of the accepted events, each of which consists of a pair of hits at the Si- and CdTe-DSDs that satisfy the Compton equation (see text for detail).
    The events that include fluorescence K emission lines are filtered out.
    The blue regions in the left panel and the hatched regions in the right panel correspond to the energy window for $^{111}$In.
    }
    \label{fig:spectrum}
\end{figure*}

After filtering by energy thresholds, we select the events which have two hits, one at the Si-DSD and the other at the CdTe-DSD.
This was applied to both experimental and simulation data.
The energy scatter plot between the Si- and CdTe-DSDs for $^{111}$In data is shown on the left in Fig.~\ref{fig:spectrum}.

We then select an event comprising of a hit from the Si-DSD and the CdTe-DSD. This definition of an event was also applied to simulated data. 
The scatter plot of the energy between Si- and CdTe-DSD for  $^{111}$In is shown on the left in Fig.~\ref{fig:spectrum}.
The green shaded regions are where the events and fluorescence K emissions following the decay of $^{111}$In are found, whose energy in Si-DSD is 23\--27~keV. 
However, we are able to remove the fluorescence K events during data processing.
Then, we extract the events in which the hits in a Si-DSD and a CdTe-DSD are emerged due to a scattering and an absorption.
We check that the measured energies satisfy the Compton equation,
\begin{equation}
    \cos\theta = 1 - m_{e}c^{2} \left(\frac{1}{\gamma\prime} - \frac{1}{\gamma}\right),
\end{equation}
where $\gamma$ is the sum of the detected energies and $\gamma\prime$ is the energy detected from the CdTe-DSD;
$m_e$ and $c$ are the electron rest mass and the light velocity, respectively.
We excluded the events when $|\cos\theta|>1$ and interpret that the hit in the Si-DSD corresponds to the scattering interaction and the hit in the CdTe-DSD corresponds to the absorption.
This is valid for gamma-rays with an energy below $m_{e}c^{2}/2$.
Using these events, we obtained the spectrum of the energy sum in the Si-DSD and the CdTe-DSD (right panel of Fig.~\ref{fig:spectrum}).
Note that some of the events that the scattering for gamma-rays with an energy of 350~keV take place in the CdTe-DSD in which could also satisfy the condition of $|\cos\theta|>1$.
However, we confirmed that the number of these events is less than 1\% of the total events, and they are negligible in this analysis.

The energy windows were set to 168--174~keV and 242--248~keV for the two primary lines of $^{111}$In from which the energy sum was calculated to analyses the ARM distribution and in the image reconstruction process.
The semiconductor detectors have sufficiently high energy resolutions so that the events consisting of summed peaks at the 194~keV and 268~keV can be completely distinguished from 171~keV and 245~keV peaks.
The energy windows for the other sources were set to 362--368 keV for $^{131}$I, 353--359 keV for $^{133}$Ba, and 508--514 keV for $^{22}$Na.

\begin{figure}[htb]
    \centering
    \includegraphics[width=0.9\linewidth]{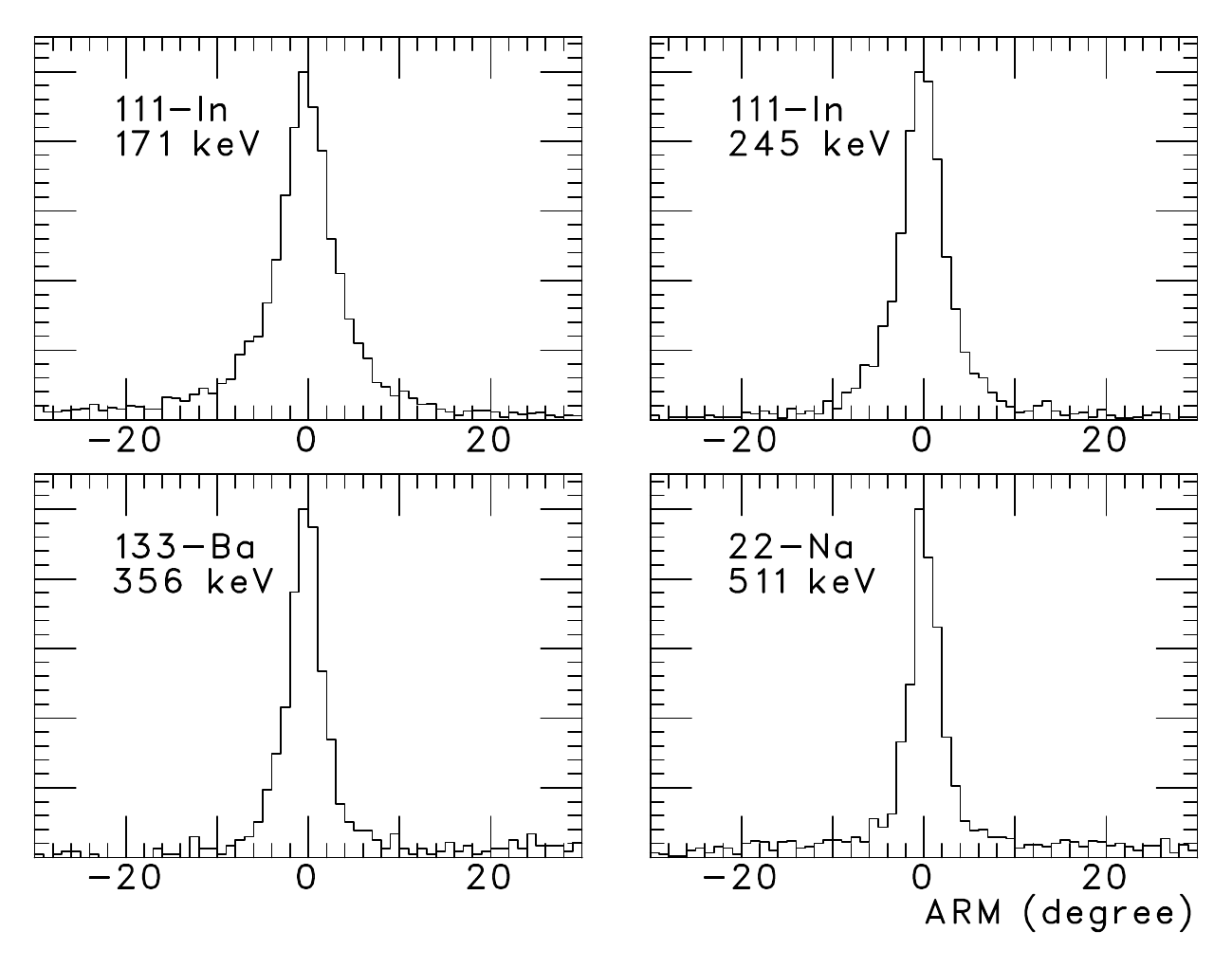}
    \caption
    {
    Distributions of the ARM obtained from our experimental data.
    The angular resolutions are 6.1$^{\circ}$ at 171~keV, 5.0$^{\circ}$ at 245~keV, 4.0$^{\circ}$at 356~keV, and 3.8$^{\circ}$at 511~keV.
    }
    \label{fig:arm4}
\end{figure}

The ARM for certain energy bands was evaluated by calculating the difference between the two scattering angles; $\theta_{K}$ was calculated from the energies of Compton scattering and photo absorption, and $\theta_{G}$ from the positions of the two hits and the source.
Fig.~\ref{fig:arm4} shows the ARM distributions at the peak energies, the angular resolution at each energy was determined at FWHM.

\subsection{Estimation using Monte-Carlo simulations}

In this section, we study how the detector parameters, \textit{i.e.}, the position and energy resolutions, affect the angular resolution of the Si/CdTe CC.
We developed the Geant4 simulation codes to simulate our experimental setting with an isotropically-emitting gamma-ray source placed at similar distances as in our experiments.
In the simulations, the two effects caused by the finite resolution in the position and energy are treated separately;
\begin{enumerate}
    \item[(a)]
    For the effect due to position resolution, we define mesh grids with the same pitch as the detector strip in a scattering volume and use the center position of the corresponding pixels instead of the precise positions of the hits. In this setup, we use a stacking pitch of 4 mm as the distance between the scattering and absorbing layers.
    \item[(b)]
    For the effect due to energy resolution, we model the detector's energy response 
    using the experimental data and apply this model to the simulation data. The detail of the model function is described in Appendix.
\end{enumerate}
Since the distance between the source and the camera is short, the size of the source is not negligible and its non-zero size has an effect on the angular resolution.
In the simulations, we set the size of the source to the same as that used in the experiments (1~mm radius).
We also modeled the geometry of the 2 Si-DSDs, 3 CdTe-DSDs plates, the aluminum detector box, and detector boards modeled using SiO$_{2}$.
The simulation records the position and energy information due to Compton scattering and photo absorption.

\begin{figure}[htb]
    \centering
    \begin{tabular}{c}
        \includegraphics[width=0.95\linewidth]{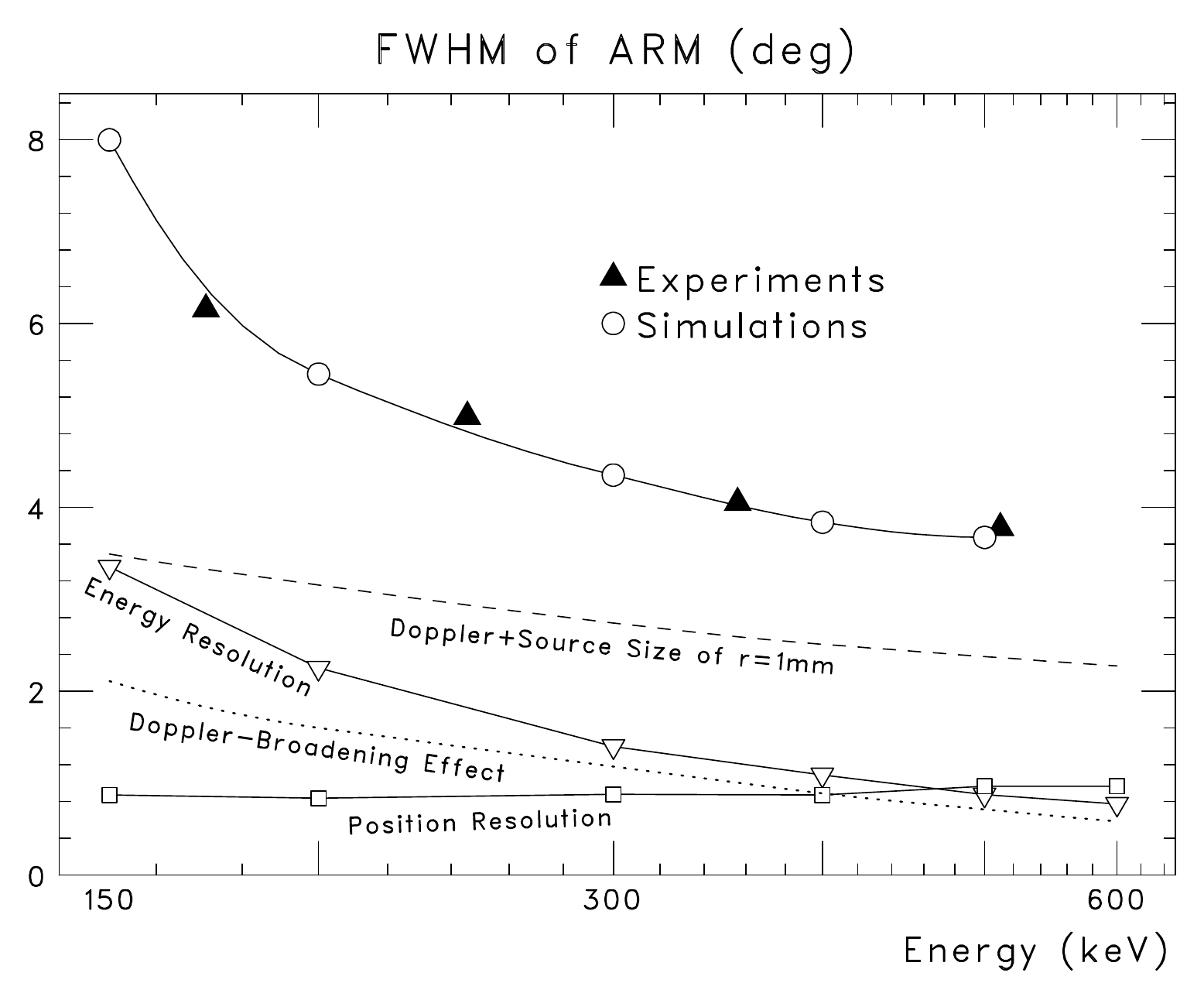}
    \end{tabular}
    \caption
    {
    Energy dependence of the angular resolution of the Si/CdTe CC measured with (filled triangles) experiments and (open circles and solid line) simulations.
    Open squares and triangles show the contributions of the energy and positional resolutions, respectively, obtained with simulations.
    Dotted and dashed lines show the FWHM of the simulated ARM distribution for a spherical source with the infinitely small radius and 1-mm radius, respectively, obtained with the ideal detector having the infinitely good positional and energy resolutions, where only the initial electron momentum distribution is considered.
    See text for detail.
    }
    \label{fig:arm}
\end{figure}

The energy dependence of the angular resolution  was obtained for both the experimental and simulation data as shown in Fig.~\ref{fig:arm}.
In the simulations, we assumed the same geometry as that of the experiment and calculated the ARMs (case 1).
We also simulated (case 2) where only the Doppler-broadening effect was considered for the detector with the infinitely good energy and positional resolutions and an infinitely small point source was assumed (dotted curve in Fig.~\ref{fig:arm}), (3) same as (2) but with a source with a size of 1~mm in radius (dashed curve), (4) where no Doppler-broadening effect, only the position resolution of the detector was taken into account (solid curve with square markers). 
(5) is similar to (4) in the energy resolution, instead of position, was taken into account (solid curve with downward-pointing open triangle markers).

The result of the main simulation (i.e., case 1) was found to be in good agreement with experiments.
We also found that using a scattering material with a small Doppler-broadening effect and a high energy resolution is important, because it significantly affects the angular resolution below 300~keV.
In addition, the simulation result (3) implies that the effect of the source size is also a significant factor for short-distance imaging.
The effect of the positional resolution is nearly constant with respect to gamma-ray energy, although a slight increase as a function of the energy is observed.
This is presumably because the uncertainty in the position of the Compton-cone axis is larger in forward scattering.
If the positional resolution was 2~mm, as an example, instead of 250~$\mu$m, this effect would be multiplied by six fold.

\subsection{Spatial resolution on the target plane}\label{sec:backp}

The spatial resolution of the Compton camera, which is a key parameter of the performance of a camera, was evaluated using the following reconstruction method.
We define the spatial resolution of the Si/CdTe CC as the FWHM of the point spread function (PSF).
The PSF is a function of the radial distance $r$ from the point source and is calculated as the integrated pixel value per area for the annulus region from $r$ to $r+dr$, where $dr$ is 0.5~mm.

We reconstructed the image by superposing multiple Compton cones onto a 2-dimensional plane at the position of the target source, parallel to the detector plane.
We refer to this method as the simple back-projection (SBP).
Specifically, we used the method described in \cite{Rohe1997, Takeda2012b} to deal with a non-zero thickness of the cone surface due to the limited position and  energy resolutions.
The pixel value at $\vec{X}$ is given by
\begin{equation}
    V( \vec{X} ) = |\vec{L}|^{-2}\exp{ \left[-\frac{1}{2} \left( \frac{x}{\sigma} \right)^2 \right] },
    \label{eq:voxel}
\end{equation}
where
\begin{equation}
    \sigma = |\vec{L}|\tan{(\Delta\theta)}
    \label{eq:sigma}
\end{equation}
and $\vec{L}$ is the closest position on the cone surface to the pixel position $\vec{X}$, $x$ is $|\vec{X}-\vec{L}|$, which is the distance between $\vec{X}$ and $\vec{L}$, and $\Delta\theta$ is the smoothing factor in the back-projection image.
In principle, $\Delta\theta$ can be negligibly small if the number of events is large enough, but in practice a very small value of $\Delta\theta$ often results in artifacts in the image.
In this work, we set $\Delta\theta$ to 0.5 degrees, the smallest value with which significant artifacts do not appear.
In order to shorten the computation time, the pixels whose distances from the cone surface were larger than 3$\sigma$ were ignored.

\begin{figure}[htb]
    \centering
    \includegraphics[width=1.0\linewidth]{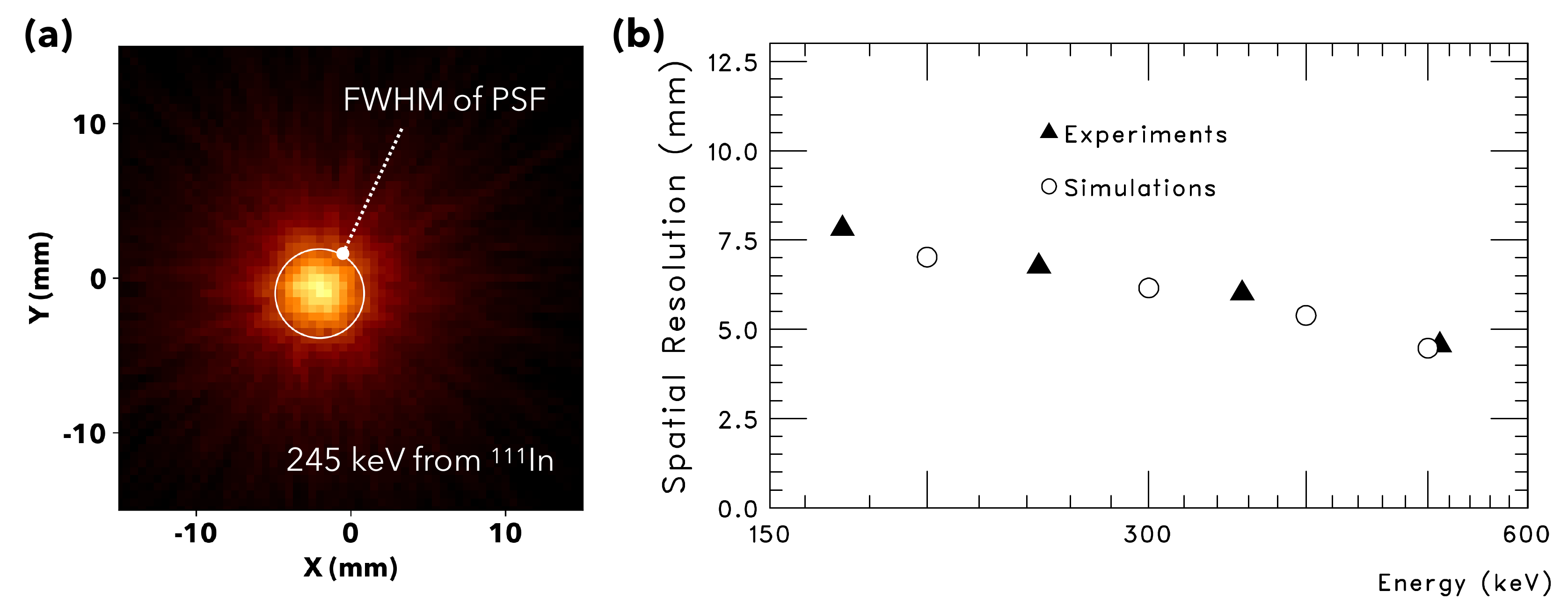}
    \caption
    {
    (a) Image constructed from experimental data with the simple back-projection for 245-keV gamma rays emitted from $^{111}$In.
    The  open circle indicates the FWHM of the point spread function.
    (b) Spatial resolution vs incident energy of the gamma rays.
    }
    \label{fig:sourcesize}
\end{figure}

Fig.~\ref{fig:sourcesize}(a) shows an image of 245~keV from $^{111}$In made with the SBP method, overlaid with the circle to indicate the FWHM of the PSF.
Fig.~\ref{fig:sourcesize}(b) shows the energy dependence of the PSF obtained from the experimental data.
The measured FWHMs of the PSF were 7.8, 6.8, and 6.0~mm at 171, 245, and 356~keV, respectively.
We confirmed that the obtained results were consistent with those of the simulations.

Hereafter, we use the FWHM of the PSF as the measure of the imaging capability rather than that of the ARM because the former is directly derived from the image.

\section{Tomographic Imaging}\label{sec:design}

\subsection{Experiment with a 3-dimensional phantom}

\begin{figure}[htb]
    \centering
    \includegraphics[width=0.85\linewidth]{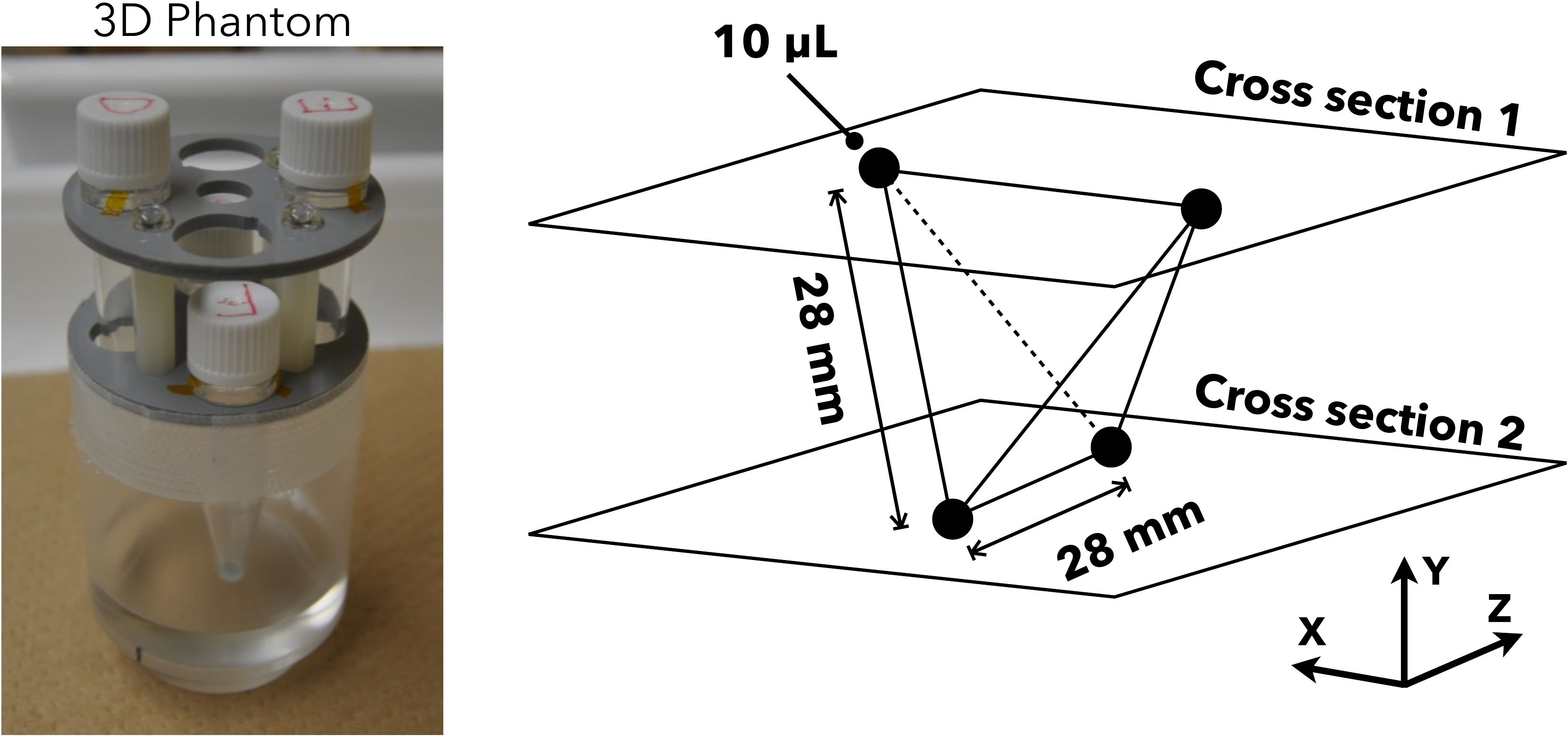}
    \caption
    {
    The phantom: each point source consists of a 10-$\mu$L droplet of liquid radionuclide, which is the light blue point at the bottom of the micro tube in the photo and is located at the vertices of a tetrahedron.
    Cross sections 1 and 2 are slices chosen for Figs.~\ref{fig:image111_In} and \ref{fig:image131_I}.
    The direction of the $y$-axis is the same as that in Fig.\ref{fig:setup}.
    }
    \label{fig:phantom}
\end{figure}

\IEEEPARstart{T}{o} evaluate the 3-dimensional imaging capability of a Si/CdTe CC, we prepared a phantom composed of 4 point sources arranged in a regular tetrahedron with side lengths of 28~mm (Fig.~\ref{fig:phantom}).
Each point source in the phantom consists of a 10~$\mu$L droplet of liquid radiation source with activities of 74~kBq held in a micro tube. 
The diameter of the droplet is estimated to be 2.7~mm.
The 4 sources had the same emission intensities within 0.2~$\%$.
The radionuclides used in the experiments were $^{111}$In and $^{131}$I, both of which are widely utilized in nuclear medicine.

We measured the phantom from 16 angles, rotating the phantom  about the vertical $y-$axis by 22.5~degree increments 
in order to acquire a 3-dimensional image, refer to Fig.~\ref{fig:setup}.
The center of the phantom was placed at a short distance of 41~mm from the camera.
The measurement time was set to 20~minutes for each  measurement.

\subsection{Reconstruction of a 3-dimensional image}

In this section, we reconstruct 3D images using the SBP method and an iterative estimation method.

Using the SBP method, a back-projection of Compton cones to a 3-dimensional imaging space is required.
We used the back-projection method described in Section~\ref{sec:backp} to calculate the voxel values of 3-dimensional histograms and superimposed the multiple back-projected Compton cones on the 3-dimensional space of the Si/CdTe CC.

We converted the positional information of the data to offset the rotation angle of the phantom.
This conversion makes the measurement of the phantom rotating with a pitch of 22.5~degrees in the experiment to be equivalent to measuring the fixed phantom from 16 angles of views.

In the simple back-projection method, the resulting image was blurred by the detector response, which was expected given that it is known to be difficult to estimate the actual distribution of gamma-ray sources using this method \cite{Takeda2009}.
Thus, we also utilized the Maximum Likelihood Expectation Maximization (MLEM) method.
The MLEM algorithm is an iterative method to find the source distribution that yields the highest likelihood from the data and it can recover the shape and size of the gamma-ray emitting sources.

Specifically, we adopted the list-mode MLEM (LM-MLEM) method described in \cite{Wilderman1998,Harrison2004}.
In this method, the image $\lambda^{(k+1)}_{j}$ after $k+1$-th iteration is calculated  from the $k$-th image $\lambda^{(k)}_{j}$ according to the formula
\begin{eqnarray} \label{eq:LM}
    \lambda^{(k+1)}_{j} &=& \frac{ \lambda^{(k)}_{j} }{ s_{j} } \sum_{i} \frac{ t_{ij} }{ \sum_{k}t_{ik}\lambda_{k} },
\end{eqnarray}
where $j$ represents the index of the voxel in the image space, $s_{j}$ is the probability that a photon emitted from the voxel $j$ can be detected and $t_{ij}$ is the probability that a gamma-ray emitted from  the $j$-th voxel will be detected as  event $i$. In this work, we use Eqs (\ref{eq:voxel}) and (\ref{eq:sigma}) to calculate this quantity.
In this case, $\Delta\theta$ in Eq (\ref{eq:sigma}) should be the angular resolution of the Compton camera, and we set it to 2.0 degrees as a representative number taken from Fig. \ref{eq:sigma}.
In this study, we focused on the size and  shape of the source reconstructed with the MLEM method, and simply assumed that $s_{j} = 1$ for all $j$.
This is because the absolute value of $s_{j}$ simply determines the normalization of the reconstructed image and does not affect the shape or the size of the reconstructed image.
Here we also assumed that all  $s_{j}$ have the same value, i.e., the detection probability was uniform across the image space.

\begin{figure}[htb]
    \centering
    \includegraphics[width=0.4\linewidth]{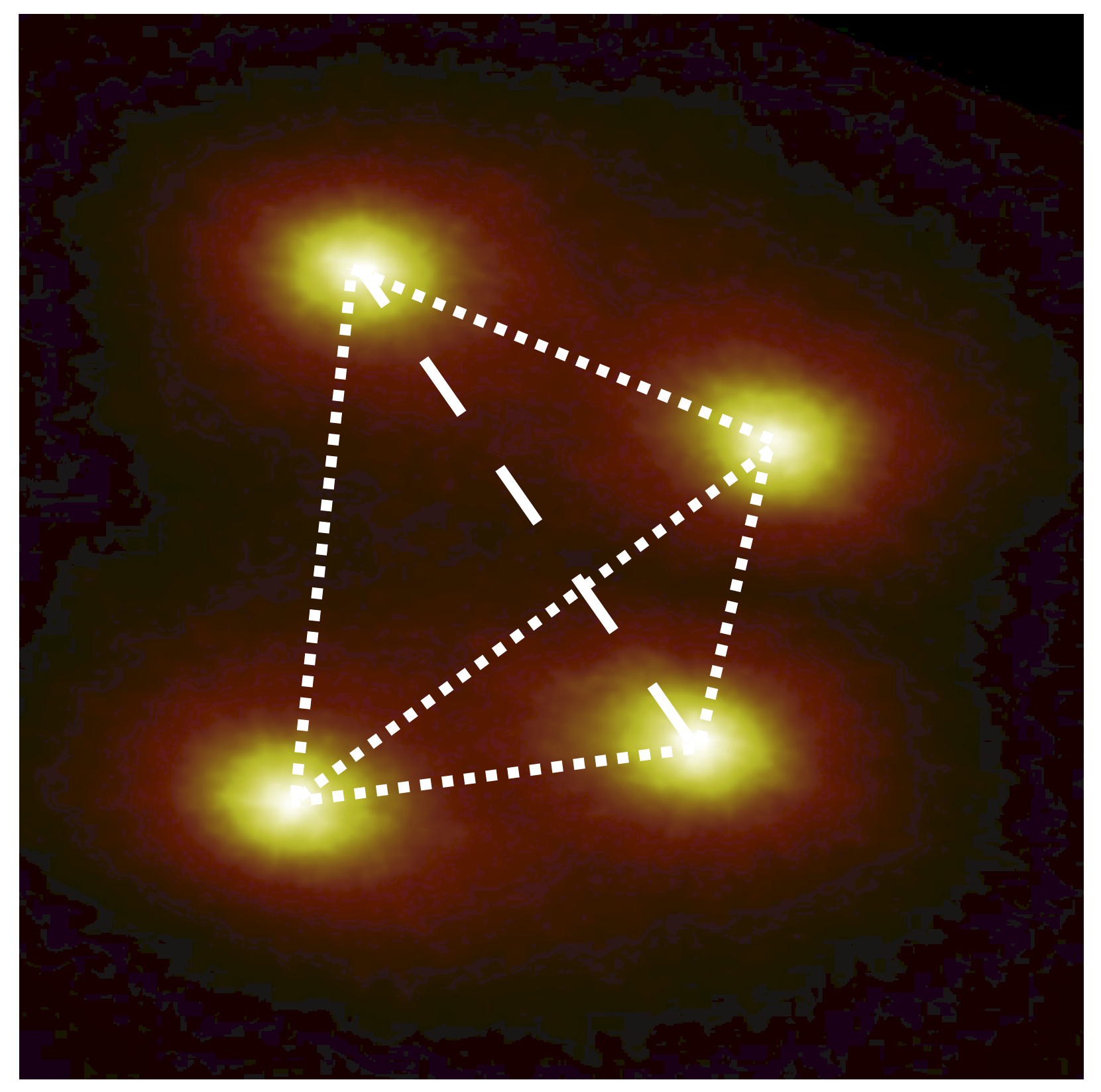}
    \caption
    {
    Simple back-projection image of the phantom visualized with the maximum intensity projection (MIP).
    This image is reconstructed from 8 projection angles of $^{131}$I data.
    }
    \label{fig:mip}
\end{figure}

\begin{table}[htb]
    \centering
    \caption{Number of the events selected for the image reconstructions}

    \begin{tabular}{ccccc} \hline
        \multicolumn{1}{c}{\# of views} & 2 & 4 & 8 & 16 \\
        \multicolumn{1}{c}{(Pitch of angles)} & (180$^\circ$) & (90$^\circ$) & (45$^\circ$) & (22.5$^\circ$) \\ \hline \hline
        \multirow{3}{*}{\begin{tabular}[c]{@{}l@{}}$^{111}$In\\ $^{131}$I\end{tabular}} &
        \multirow{3}{*}{\begin{tabular}[c]{@{}l@{}}36056\\ 12764\end{tabular}} &
        \multirow{3}{*}{\begin{tabular}[c]{@{}l@{}}71038\\ 24502\end{tabular}} &
        \multirow{3}{*}{\begin{tabular}[c]{@{}l@{}}141585\\ 47481\end{tabular}} &
        \multirow{3}{*}{\begin{tabular}[c]{@{}l@{}}284790\\ 94276\end{tabular}} \\
        & & & & \\
        & & & & \\ \hline
    \end{tabular}

    \label{tab:numberofevents}
\end{table}

\section{Results and discussion}\label{sec:result}

\IEEEPARstart{W}{e} reconstructed a simple back-projection image using the events extracted  with the selection criteria  described in Section~\ref{sec:selection}.
Figure \ref{fig:mip} shows the reconstructed SBP image of a $^{131}$I phantom measured from 8 view angles, using the maximum intensity projection (MIP) method for visualization.
The numbers of events selected from each data set for use in image reconstructions are listed in Table~\ref{tab:numberofevents}.

\begin{figure}[thb]
    \centering
    \includegraphics[width=0.9\linewidth]{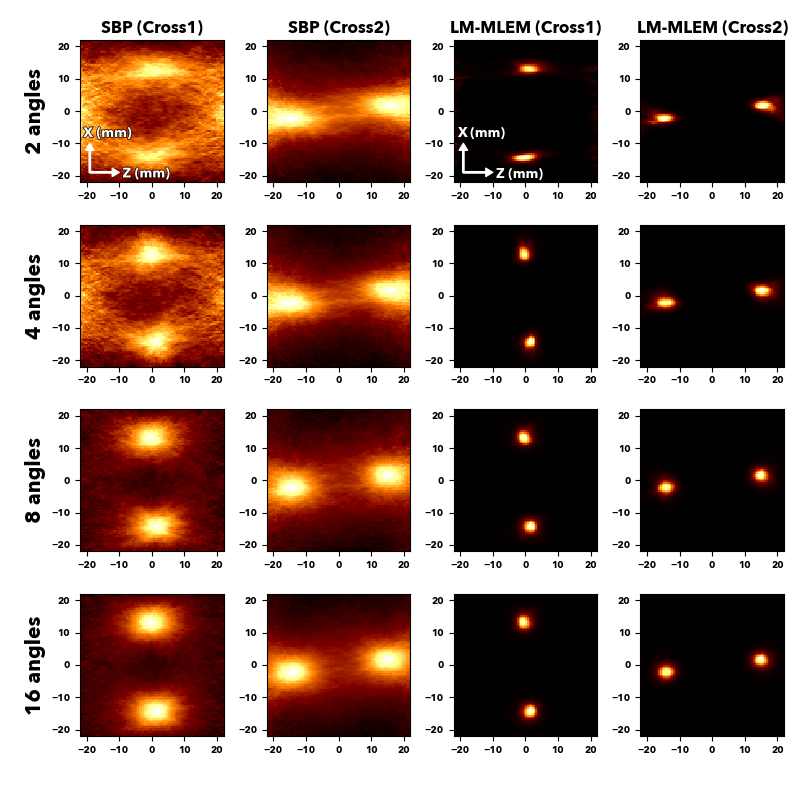}
    \caption
    {
    SBP images and LM-MLEM images for the $^{111}$In phantom sliced at the Cross-sections 1 and 2 (see Fig~\ref{fig:phantom}).
    Images on each rows are reconstructed from the data measured from 2, 4, 8, and 16~angles.
    }
    \label{fig:image111_In}
\end{figure}

\begin{figure}[thb]
    \centering
    \includegraphics[width=0.9\linewidth]{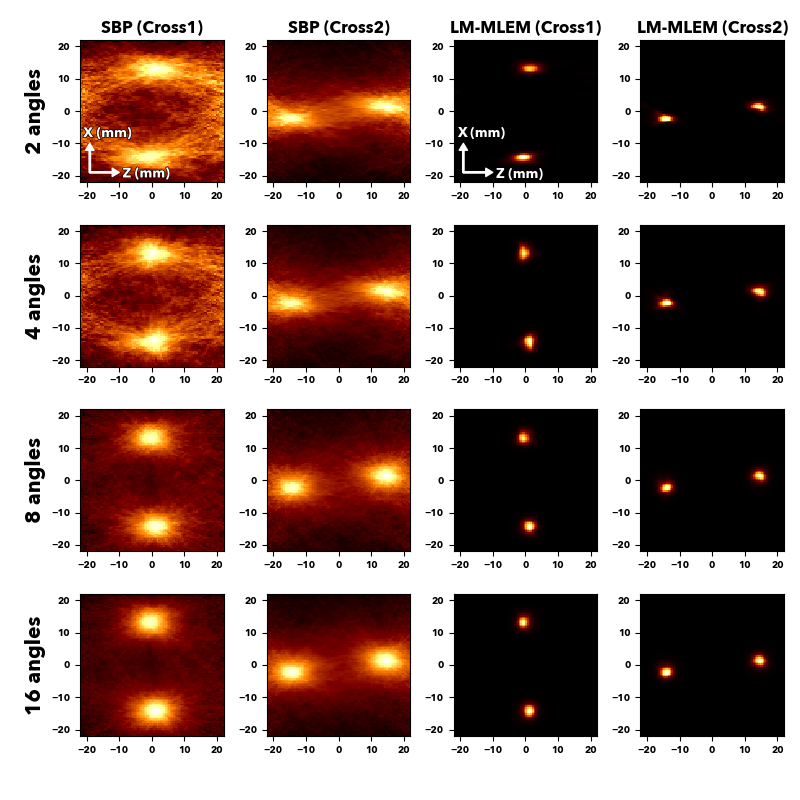}
    \caption
    {
    Same as Fig.\ref{fig:image111_In} but those for the $^{131}$I phantom.
    }
    \label{fig:image131_I}
\end{figure}

The 3-dimensional structure of the phantom, which is a tetrahedron (see Fig.\ref{fig:mip}), was successfully reproduced, and the four point sources were clearly visible.
The position of each point was determined as the center of the voxel, which is the position of the local maximum values, and the distances between them were found to be consistent with the expected configuration with deviations away from the true value of less than 1.5~mm.
The spatial resolution of the 3-dimensional image was evaluated by means of  the FWHM of a point in cross-sectional slices of the back-projection images.
The panels in the left 2 columns in Figs.~\ref{fig:image111_In} and \ref{fig:image131_I} show lateral cross-sectional slices (see Figs.~\ref{fig:setup} and \ref{fig:phantom}) for $^{111}$In and $^{131}$I, respectively, which were reconstructed from the data measured from 2 (top row), 4, 8, and 16 angles.
The lateral resolution was measured with the cross-sectional slices parallel to the $x-z$ plane in Figs.~\ref{fig:setup} and \ref{fig:phantom}. 
We measured the FWHMs of the distribution of the voxel values along the $x-$ and $z-$ axes of the sliced images of point sources of the two radionuclides.
The larger and smaller ones are described as the lateral resolutions along the major and minor axis, respectively, in Table~\ref{tab:pointsize} and \ref{tab:pointsize2}.
The axial resolution was defined as the FWHM of the distribution of the voxel values along the $y$-axis.

For the images reconstructed from 16 projection angles, we obtained lateral resolutions of 11.5 and 9.0~mm for $^{111}$In and $^{131}$I, respectively.
The axial resolutions were 6.1 and 4.4~mm, respectively.
Table~\ref{tab:pointsize} summarizes the result.

\begin{table}[thb]
    \centering
    \caption
    {
    Spatial resolution  of the SBP images
    }
    \begin{tabular}{|c|c|c|c|}
        \multicolumn{4}{c}{Resolution based on images for $^{111}$In (171~keV+245~keV)} \\ \hline

        \multirow{5}{*}{\# of angles} & \multicolumn{2}{c|}{\multirow{2}{*}{Lateral resolution}}  & \multirow{5}{*}{Axial resolution} \\
        & \multicolumn{2}{c|}{} &  \\ \cline{2-3}
        &  \multirow{2}{*}{Major  axis} & \multirow{2}{*}{Minor  axis}  & \multirow{2}{*}{} \\
        &  \multirow{2}{*}{[mm]} & \multirow{2}{*}{[mm]}  & \multirow{2}{*}{[mm]} \\
        & & & \\ \hline \hline
        2   & 20.3 &  8.1 &  7.7  \\
        4   & 17.6 &  8.1 &  6.8  \\
        8   & 11.4 &  7.6 &  6.1  \\
        16  & 11.5 &  7.2 &  6.1  \\
        \hline
        \multicolumn{4}{c}{} \\
        \multicolumn{4}{c}{Resolution based on images for $^{131}$I (364~keV)} \\ \hline

        \multirow{5}{*}{\# of angles} & \multicolumn{2}{c|}{\multirow{2}{*}{Lateral resolution}}  & \multirow{5}{*}{Axial resolution} \\
        & \multicolumn{2}{c|}{} &  \\ \cline{2-3}
        &  \multirow{2}{*}{Major  axis} & \multirow{2}{*}{Minor  axis}  & \multirow{2}{*}{} \\
        &  \multirow{2}{*}{[mm]} & \multirow{2}{*}{[mm]}  & \multirow{2}{*}{[mm]} \\
        & & & \\ \hline \hline
        2   & 15.6 &  5.2 &  5.0  \\
        4   & 13.1 &  5.8 &  4.9  \\
        8   &  8.7 &  6.0 &  4.2  \\
        16  &  9.0 &  5.9 &  4.4  \\
        \hline

    \end{tabular}\\
    \vspace{5pt}
    The lateral resolution columns show the FWHM of the point source along the $x-$ and $z-$axes, whereas the axial resolution one shows that along the $y-$axis.
    See Fig.\ref{fig:phantom} for the definition of the axes.
    \label{tab:pointsize}
\end{table}

\begin{table}[thb]
    \centering
    \caption
    {
    Spatial resolutions of the LM-MLEM images
    }
    \begin{tabular}{|c|c|c|c|}
        \multicolumn{4}{c}{Resolution based on images for $^{111}$In (171~keV+245~keV)} \\ \hline
        \multirow{5}{*}{\# of angles} & \multicolumn{2}{c|}{\multirow{2}{*}{Lateral resolution}}  & \multirow{5}{*}{Axial resolution} \\
        & \multicolumn{2}{c|}{} &  \\ \cline{2-3}
        &  \multirow{2}{*}{Major width} & \multirow{2}{*}{Minor width}  & \multirow{2}{*}{} \\
        &  \multirow{2}{*}{[mm]} & \multirow{2}{*}{[mm]}  & \multirow{2}{*}{[mm]} \\
        & & & \\ \hline \hline
        2    &  4.1 &  1.8 &  2.0  \\
        4    &  4.6 &  2.6 &  2.5  \\
        8    &  4.1 &  3.0 &  2.7  \\
        16   &  4.0 &  3.1 &  2.7  \\
        \hline
        \multicolumn{4}{c}{} \\
        \multicolumn{4}{c}{Resolution based on images for $^{131}$I (364~keV)} \\ \hline

        \multirow{5}{*}{\# of angles} & \multicolumn{2}{c|}{\multirow{2}{*}{Lateral resolution}}  & \multirow{5}{*}{Axial resolution} \\
        & \multicolumn{2}{c|}{} &  \\ \cline{2-3}
        &  \multirow{2}{*}{Major width} & \multirow{2}{*}{Minor width}  & \multirow{2}{*}{} \\
        &  \multirow{2}{*}{[mm]} & \multirow{2}{*}{[mm]}  & \multirow{2}{*}{[mm]} \\
        & & & \\ \hline \hline
        2    &  3.1 &  1.5 &  1.8  \\
        4    &  3.2 &  2.0 &  2.1  \\
        8    &  2.8 &  2.2 &  2.1  \\
        16   &  2.7 &  2.4 &  2.1  \\
        \hline
    \end{tabular}\\
    \vspace{5pt}
    See Table~\ref{tab:pointsize} caption for the detailed description.
    \label{tab:pointsize2}
\end{table}

The shape of each point source was somewhat distorted to an oval shape from the actual circular (or spherical in the three dimension) shape when the number of the observed angles was small, and the tendency was more pronounced with a smaller number of angles (Figs.~\ref{fig:image111_In} and \ref{fig:image131_I}).
The distortion in the images is reflected in Table~\ref{tab:pointsize} as a large difference between the major- and minor-axis widths of a point source; the former is about 2.5 times larger than the latter for the 2-angle images.
For the image of 8 angles, the difference between the major- and minor-axis widths is smaller than these widths.
There is a tendency for the difference of the width to be smaller for increasing as the number of the observation angles increases.
Given that the improvement of the width differences from the images of 8 angles to those of 16 angles is much smaller than the widths, observing in more than 8 projection angles does not improve the image quality very much.
In the SBP image, the lateral resolutions were significantly worse than the axial ones; this is consistent with the fact that the region, created by superimposing multiple Compton cones, extends in the $x-z$ plane for the present configuration of the camera.
The obtained image is consistent with this view.
Here, the $y-$axis is defined to be perpendicular to the plane of the rotation of the phantom ($x-z$ plane in Fig.\ref{fig:phantom}).


We reconstructed the image using the LM-MLEM method.
The calculation was stopped after 20 iterations for each dataset because any more iteration after the 20-th iteration would not change the derived size of each point source in the reconstructed image by more than 1\%.

Using the LM-MLEM method, the sizes of the point sources converged to those of the actual phantom in the 3-dimensional space, and the source shapes were also reproduced.
The panels in the right 2 columns of Figs.~\ref{fig:image111_In} and \ref{fig:image131_I} show the lateral cross-sectional slices of the images.
The oval distortion for the 4-angle observation in the LM-MLEM images appeared less significant than that of the corresponding simple back-projection images (second rows in Figs.~\ref{fig:image111_In} and \ref{fig:image131_I}).
For images from 8 (third rows) and 16 (fourth rows) projection angles, distortions of the point sources were not visible anymore.
Table~\ref{tab:pointsize2} tabulates the results.
We obtained lateral resolutions to be 4.0 and 2.7~mm for the radiation sources $^{111}$In and $^{131}$I, respectively, for the images reconstructed from 16 projection angles.
The axial resolutions were 2.7 and 2.1~mm, respectively.

In the image of the 8-angles and 16-angles, the spatial resolution ranges from 2.1--4.1~mm for gamma rays below 400~keV, the energy of which is low for imaging with Compton cameras.
From the symmetry of the actual source shape, its major and minor widths are expected to be the same.
We found that the difference between the major/minor widths was smaller in the MLEM than in the SBP method, the fact of which implies that the MLEM reproduces the true shape of the source better.
It should be noted that the spatial resolution described here includes the source size. Therefore, our results imply that the Si/CdTe Compton camera has better performances in localization of sources.

We also calculated the source intensities of the reconstructed images.
Specifically, we evaluated the uniformity of the intensity with the counts in ROIs (region of interest) set on each point source in the reconstructed images from 16 projection angles.
The sizes of these ROIs are a sphere with a 10~mm diameter, which is large enough to cover the distributions of the point sources.
The deviations from the averaged intensity the 4 point sources were calculated to be around 10\% for both $^{111}$In and $^{131}$I.
Similar results were obtained from the images of the LM-MLEM method.

In summary, we demonstrated that the Si/CdTe Compton camera gives the spatial resolution of 4.4-11.5~mm at a distance of 41~mm from the detector surface in the SBP image.
The localization in the tetrahedron-shaped phantom is much improved when we use the LM-MLEM method.
The resultant size of the sources in the reconstructed image ranges from 2.1--4.1~mm, compared to the actual size of the droplet of the radioactive sources of 2.7~mm (Fig.~\ref{fig:sourcesize}).
This performance is attractive for the use in practical applications of imaging of gamma-rays by using commonly-used radionuclides, such as $^{111}$In and $^{131}$I.

\section{Conclusions}

\IEEEPARstart{W}{e} conducted a study of imaging capability of a Si/CdTe CC using  3-dimensional phantoms with $^{111}$In and $^{131}$I solutions.
These radionuclides, which are often used in medical imaging, emit low-energy gamma rays below 0.5~MeV.
Conventional Compton cameras are not designed to cover such a low energy band with a spatial resolution better than 5~mm.
The Si/CdTe CC has a good energy and positional resolutions and its Doppler-broadening effect is smaller than traditional Compton cameras based on scintillators such as NaI(Tl).
We demonstrated that tomographic imaging of a Si/CdTe CC yields improved image quality.
The SBP and LM-MLEM algorithms were applied to a set of projection data obtained from 16 different angles.
Using simple back-projection methods, the positions of the sources are reproduced with a lateral resolution of 11.5~mm and 9.0~mm (FWHM) for $^{111}$In and $^{131}$I, respectively.
We found that a LM-MLEM method gives better lateral resolutions of 4.0~mm and 2.7~mm (FWHM) with 16 view angles.
These numbers are reduced to 2.7~mm and 2.1~mm, when we use the axial resolution.
We successfully separated the sources which are placed at a distance of 28~mm in the tetrahedron structure of the phantom.
The results imply the potential of Compton Camera in the energy range below 400~keV for observing targets placed close to the camera.

\appendix
\subsection{Estimation of the Doppler-broadening effects of silicon}
\label{appen:doppler}

\begin{figure}[htb]
    \centering
    \includegraphics[width=0.9\linewidth]{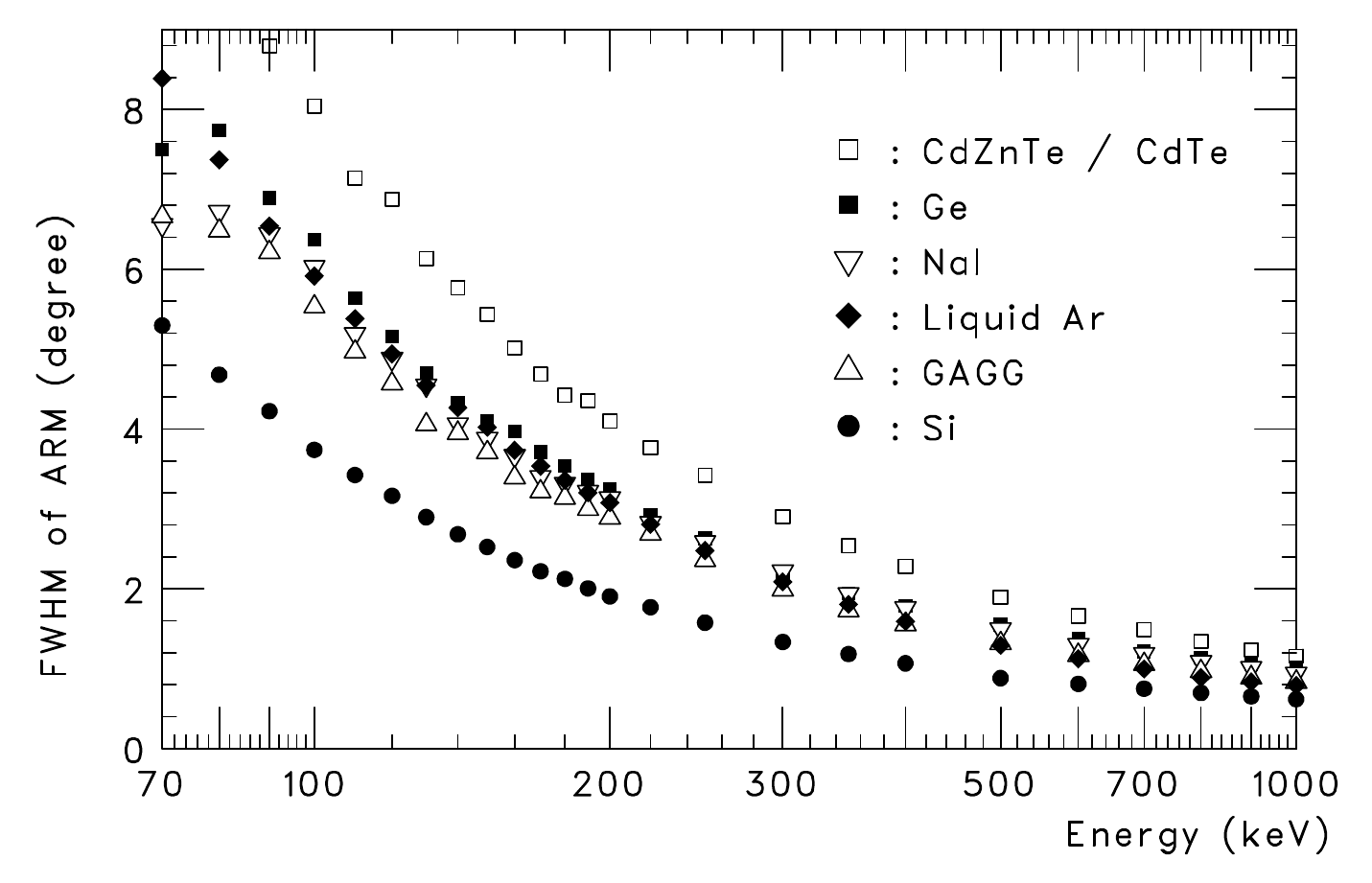}
    \caption
    {
    Contributions of the Doppler broadening effect to the angular resolution of Compton cameras with different materials for the scattering layers as a function of the energy, simulated with the Geant4 physics model G4EmLivermorePhysics (see text).
    The vertical axis is the FWHM of the ARM distributions, which is interpreted as the angular resolution of the camera.
    }
    \label{fig:doppler}
\end{figure}

We present our study of the Doppler-broadening effects of silicon and other scatter materials in the detector.
The effect is represented as the angular resolution in this section.
We calculated the angular resolution, using the Geant4\cite{Geant4-2003} simulation toolkit, where we assumed a 100~m$\times$100~m$\times$100~m cube for each material and recorded the energies and momenta when gamma rays hit the material (Fig.~\ref{fig:doppler}).
Here, again the angular resolution of the Compton camera is defined as the FWHM of ARM.
Silicon was found to show the best angular resolution among the materials, which were about 50 and 70~\% of those of CdZnTe and GAGG, respectively.
Given the lowest Doppler-broadening effect among the candidate materials, we conclude that  silicon is the most suitable scattering material.

\subsection{Energy resolutions of the detectors in simulations}
\label{appen:deltae}

For the simulations used in this study, the energy resolution of the detectors is determined   as follows.
The energy resolution ($\Delta E$) of the semiconductor detector is modeled with the sum of electronics noise ($\Delta E_{electronics}$) and statistical fluctuation ($\Delta E_{statistics}$) given by
\begin{equation}
    \Delta E = \sqrt{(\Delta E_{electronics})^{2}+(\Delta E_{statistics})^{2}},
\end{equation}
where
\begin{equation}
    \Delta E_{statistics} = 2.35\sqrt{FEW},
\end{equation}
and $F$ is the Fano factor, $E$ is the gamma-ray energy, and $W$ is the average ionization energy.
The Fano factors for silicon and CdTe are 0.11 and 0.13, respectively.
The average ionization energies are 3.6~eV and 4.5~eV, respectively.
$\Delta E_{electronics}$ is measured to be 1.6~keV in both Si and CdTe.
For CdTe-DSDs, a term to represent the decrease in energy resolution caused by the depth dependence of the charge collection efficiency in the detectors is also needed.
This is approximated using the following linear function
\begin{equation}
    \Delta E_{cdte} = \sqrt{(\Delta E_{electronics})^{2}+(\Delta E_{statistics})^{2}}  + \alpha E.
\end{equation}
Fitting this to the experimental data yields $\alpha$ of 0.008.

\section*{Acknowledgment}
This work was supported by JSPS KAKENHI Grant Numbers 16H02170, 16H03966, 18H02700, J18H05463, 20K22355 and World Premier International Research Center Initiative (WPI), MEXT, Japan.
This work was supported by a matching fund program of Centers for Inter-University Collaboration from ISAS (Institute of Space and Astronautical Science), JAXA (Japan Aerospace Exploration Agency).
This work was supported by JST Advanced Measurement and Analysis program (2012-2014).
The authors thank iMAGINE-X Inc. for supporting data analysis of Compton cameras.

\ifCLASSOPTIONcaptionsoff
\fi

\bibliographystyle{IEEEtran}
\bibliography{IEEEabrv,refer.bib}


\end{document}